\documentclass{aa}
\usepackage{verbatim,float}
\usepackage{times}
\usepackage{amsmath}
\usepackage{graphics}
\usepackage{epsfig}
\usepackage{psfig}

\begin{document}
\title{The Globular Cluster Systems of \object{NGC 3258} and \object{NGC 3268} in the \object{Antlia Cluster} 
	\thanks{Based on observations collected at the Cerro Tololo Inter-American Observatory (CTIO).}}
	\author{B. Dirsch \inst{1}, T. Richtler \inst{1}, L.P. Bassino \inst{2}} 
\offprints{bdirsch@cepheid.cfm.udec.cl}
\institute{Universidad de Concepci\'on, Departamento de F\'{\i}sica,
           Casilla 160-C, Concepci\'on, Chile
       \and
	  Facultad de Ciencias Astron\'omicas y Geof\'{\i}sicas,
          Universidad Nacional de La Plata,
	  Paseo del Bosque S/N, 1900-La Plata,
	  Argentina and CONICET
       }
\titlerunning{The Globular Cluster Systems of two Antlia Cluster Galaxies}

\date{Received 16 April 2003 / Accepted 6 June 2003}

\abstract{
The Antlia galaxy cluster is the third nearest galaxy cluster after Virgo and
Fornax. We used the wide-field MOSAIC camera of the 4-m CTIO telescope to
search in the brightest
cluster galaxies for globular cluster systems, which were detected in the two
larger ellipticals -- NGC\,3258 and NGC\,3268. These galaxies each contain 
several thousand clusters; NGC\,3258 more than NGC\,3268. The color distributions
of the globular cluster systems are clearly bimodal. The peak colors agree with
those of other ellipticals. The radial number density profiles of
the globular cluster systems  
are  indistinguishable for the two galaxies and no difference in the
distribution of red and blue clusters - as observed in other elliptical galaxies -
can be seen. 
The light profile of NGC\,3268 appears to be similar to that of NGC\,1399, the 
central galaxy of the Fornax cluster. NGC\,3258 has a light profile which is steeper
at large radii.
Both galaxies exhibit color gradients, becoming bluer outwards.
In NGC\,3268, the color and morphology in the inner 3\arcsec \ indicate the
presence of an inner dusty disk.
The globular cluster systems closely trace the galaxy light in the studied radial
regime. The elongation of the cluster systems of both galaxies is approximately 
aligned
at large radii with the connecting axis of the two galaxies.
We find  specific frequencies within a radial range of
4\arcmin \ of $\mathrm{S}_\mathrm{N}=3.0\pm2.0$ for NGC\,3268 and
$\mathrm{S}_\mathrm{N}=6.0\pm2.5$ for NGC\,3258.

As a byproduct resulting from surveying our wide-field frames, we describe a strange 
absorption feature in the Antlia spiral galaxy NGC 3269, which we
argue might be a tiny galactic dust cloud projected onto NGC 3269.

\keywords{Galaxies: elliptical and lenticular, cD -- Galaxies: individual: NGC 3258, NGC 3268, NGC 3269 -- 
Galaxies: star clusters -- Galaxies: stellar content -- Galaxies: structure}
}

\maketitle
\section{Introduction}

\subsection{The Antlia cluster}

The Antlia galaxy cluster, located in the Southern sky at low Galactic latitude
(19\degr), is after Virgo and Fornax the nearest galaxy cluster. It is described
as ``a beautiful, small, nearby cluster of Bautz-Morgan type III'' (Hopp \& Materne
\cite{hopp85}, Nakazawa et al. \cite{nakazawa00}). As already noted by Hopp \&
Materne (\cite{hopp85}), there exists a slight confusion in the nomenclature: the
Antlia cluster as defined by Sandage (\cite{sandage75}) should not be mistaken for
the Antlia group defined by Tully et al. (\cite{tully82}) that belongs to the Local
Super Cluster.

The galaxy content of the Antlia cluster has been studied by Hopp \& Materne
(\cite{hopp85}) and by Ferguson \& Sandage (\cite{ferguson90}, \cite{ferguson91}).  
Ferguson \& Sandage (\cite{ferguson90}) identified in total 234 Antlia cluster
galaxies, compared to 340 in Fornax. After correction for the completeness of
the sample and taking the different distances into account, they found
that the total number of Antlia cluster galaxies is larger than that of the Fornax
cluster within a radial distance of 5 times the core radius (420 versus 330
galaxies).

Of the seven nearby groups/clusters studied by Ferguson \& Sandage
(\cite{ferguson90}, \cite{ferguson91}), the Antlia cluster has the highest 
central galaxy number density (e.g. roughly three times larger than in
Fornax or Virgo, depending on the assumed distance).  In accordance with the large
density, its galaxy population is dominated by early-type galaxies.  The dwarf
ellipticals are by far the most abundant galaxy type in this cluster.

The Antlia cluster's galaxy velocity dispersion ranges from 360\,km/sec to
560\,km/sec, depending on the galaxy sample. One complication is that there might
be a connection from the Antlia cluster to the Hydra I cluster that has a higher
redshift (3400\,km/sec versus 2900\,km/sec) (Hopp\,\&\,Materne \cite{hopp85}). In
addition, it has been found that the early-type galaxies have a systematically lower
radial velocity with respect to late-type galaxies, causing some confusion in the
velocity dispersion determination (Hopp\,\&\,Materne \cite{hopp85}).

The central part of the Antlia cluster consists of two subgroups (see
Fig.\,\ref{fig:antliafield}), each dominated by a giant elliptical galaxy. The Northern
subgroup contains the early-type galaxy NGC\,3268 and three larger late-type
galaxies. The dominating elliptical of the Southern subgroup is NGC\,3258 that is
accompanied by two smaller ellipticals. The two dominating ellipticals, 
NGC\,3258 and NGC\,3268, nearly have the
same luminosity. The Antlia cluster is therefore of a different type than for
example the Virgo or Fornax clusters which are dominated by large 
central galaxies.

The X-ray properties of the Antlia cluster are peculiar. Pedersen et al.
(\cite{pedersen97}) used ASCA observations centered on NGC\,3258, where they
clearly found extended X-ray emission. They also noted that the 1.7\,keV hot gas
extends towards NGC\,3268 and they excluded this part from their investigation. 
The emission peak is offset by 1.1\arcmin \ from NGC\,3258 and does
not coincide with any other galaxy. They derived a total mass for the NGC\,3258 
group of
$0.9-2.4\times10^{13}\mathrm{M}_{\sun}$ within 240\,kpc. 
The NGC\,3268 group has been studied by  Nakazawa et al.
(\cite{nakazawa00})  using ASCA and ROSAT data both centered on NGC\,3268. They
confirmed  the X-ray emission extending towards NGC\,3258 and excluded this direction
as well.
According to their study Antlia is embedded in a quasi isothermal X-ray
halo with a temperature of 2\,keV. As in Pedersen et al. (\cite{pedersen97}), no
excess central brightness connected with the central galaxy, as frequently seen in
nearby clusters, can be detected. This distinguishes the Antlia cluster from
clusters like Fornax (Ikebe et al. \cite{ikebe96}) or Hydra-A (Ikebe et al.
\cite{ikebe97}).
Nakazawa et al. (\cite{nakazawa00}) used their X-ray observations to derive a total
gravitating mass of $1.9\times 10^{13}$M$_{\sun}$ within a radius of 250\,kpc.  
Strikingly, both groups obtained cluster masses that are of the same order of
magnitude (these masses are measured within a radius that encompasses the two
central galaxies) as the mass derived for the optically larger Fornax cluster within the
same radius (Jones et al. \cite{jones97}).

While the early-type to late-type galaxy ratio 
indicates an evolved system, the existence of two subgroups, which also may
be present in the overall mass distribution, means that
the total system has not yet completed its evolution. We might be witnessing
the merging of two small, rather evolved, compact clusters. 

Although Antlia is nearby and exhibits interesting properties, the globular
cluster systems (GCSs) of the dominating galaxies have not yet been studied. 
A GCS provides insight into the stellar populations of the host galaxies, where
the galaxy light itself has become too faint to be studied in detail, offering
single stellar populations. 
GCSs have also been used to trace the galaxies' evolution (e.g. Pritchet \& Harris 
\cite{pritchet90}, Ashman \& Zepf \cite{ashman92}, McLaughlin et al. \cite{mclaughlin93}, 
Forbes et al. \cite{forbes97}, Harris et al. \cite{harris00}, Harris \cite{harris01},
Beasley et al. \cite{beasley02}, C\^ot\'e et al. \cite{cote02}, Beasley et al. 
\cite{beasley03}).  Here we investigate the GCSs of
the two dominating ellipticals and study their basic properties, like color and
radial number distribution.  This is then compared to the light profiles of the
host galaxies and their
color properties.

The wide-field MOSAIC camera mounted in the prime focus of the CTIO 4-m telescope
is the ideal instrument for this task: the main body of the Antlia cluster fits
perfectly into one single MOSAIC frame. We used the Washington filter system, since
it has a very good metallicity resolution and it is rather powerful in separating
compact, blue background galaxies from cluster candidates (Dirsch et al.
\cite{dirsch03}).

\subsection{Distances of NGC\,3258 and NGC\,3268}

The distances towards NGC\,3258 and NGC\,3268 have been derived with SBF techniques
(Tonry et al. \cite{tonry01}) and are $\mathrm{(m-M)}=32.58\pm0.27$ and
$\mathrm{(m-M)}=32.71\pm0.25$, respectively. This agrees with the Hubble flow
distance if a Hubble constant of 75\,(km/sec)/Mpc is assumed. Prugniel et al.
(\cite{prugniel96}) obtained a distance of $\mathrm{(m-M)}=32.92$ for both galaxies
using the fundamental plane.
However, the extinctions used by Prugniel et al. are
taken from Burstein\,\&\,Heiles (\cite{burstein82}) which are lower by 0.08 and
0.14 in V for NGC\,3258 and NGC\,3268, respectively, than those of Schlegel et al.
(\cite{schlegel98}), which were used by Tonry et al..  Since we prefer the
reddening values of Schlegel et al., the distance modulus of Prugniel et al. would
be $\mathrm{(m-M)}=32.86$ for NGC\,3258 and $\mathrm{(m-M)}=32.78$ for NGC\,3268.  The
distances towards the two ellipticals are hence the same within the uncertainties
and we adopt in the following the mean of the four distance moduli of
$\mathrm{(m-M)}=32.73\pm0.25$ towards the Antlia cluster, so that angular 
distances of 1\arcsec \ and 1\arcmin \ correspond to 170\,pc and 10.2\,kpc, respectively. 
However, assuming one
distance for all Antlia cluster galaxies might be an oversimplification due to the
fact that NGC\,3271, NGC\,3269 and NGC\,3267 have very deviating radial velocities
(see Table\,1)  with respect to the other galaxies. Interestingly,
these three galaxies are located at very small projected radial distances from NGC
3268, for which the SBF technique gave a larger distance. 

\section{Observations \& Reduction}

\begin{figure}[t]
 \centerline{\resizebox{\hsize}{!}{\includegraphics{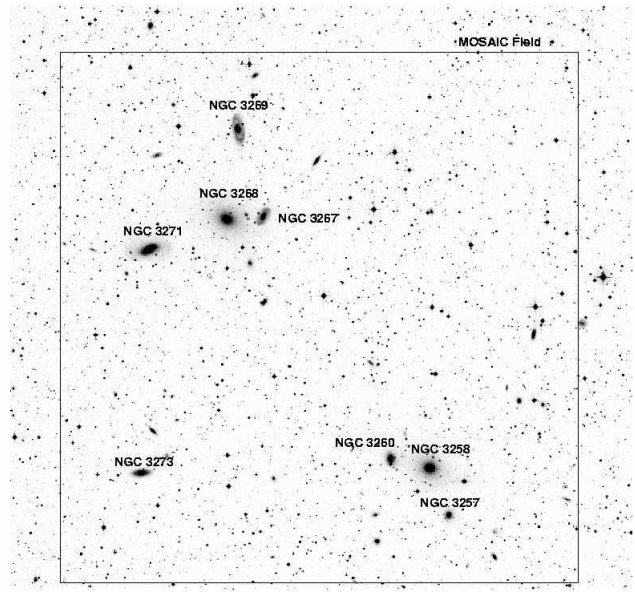}}}
 \caption{The MOSAIC field is overlaid on a DSS image of the central part of
	the Antlia cluster. North is up, East to the left.}
 \label{fig:antliafield}
\end{figure}

The data set consists of Washington wide-field images taken with the MOSAIC camera
mounted at the prime focus of the CTIO 4-m Blanco telescope during 4./5. April
2002. We obtained four 600 sec images in R and seven 900 sec images in C. In
addition, a 60 sec exposure was taken in R and a 180 sec exposure in C. The
galaxies' cores are not saturated in these short exposures. The location of the
MOSAIC field, overlaid on a DSS image, is shown in Fig.\,\ref{fig:antliafield}
with the brightest galaxies labelled.

We used the Kron-Cousins R and Washington C filters, although the genuine
Washington system uses T1 instead of R.  However, Geisler (1996) has shown that the
Kron-Cousins R filter is more efficient than T1 and that R and T1 magnitudes are
closely related, with only a very small color term and zero-point difference
(we got $\mathrm{R-T1}=-0.02$ from standard stars). The MOSAIC wide-field 
camera images a field of
$36\arcmin\times 36\arcmin$. The 8 SITe CCDs have pixel scales of
0.27\arcsec/pixel. Further information on the MOSAIC camera can be found in the
MOSAIC homepage ({\it http://www.noao.edu/kpno/mosaic/mosaic.html}).

We dithered the data to fill the gaps between the eight individual CCD chips. Due
to the dithering the entire area could not be covered by the same number of exposures which
restricts the usable field. For the final frames we thus discarded the outer regions
that showed a much higher noise than the inner area and trimmed the images to
$34\farcm7\times34\farcm7$. The seeing on the final R image is 1\arcsec \ and on the 
final C image $1\farcs1$.

The MOSAIC data has been handled using the {\it mscred} package within IRAF. In
particular, this package can correct for the variable pixel scale across the CCD
which would cause otherwise a 4\% variability of the brightness of stellar-like
objects from the center to the corners. The flatfielding residuals led to
sensitivity variations $\le 2.5\%$ (peak-to-peak).

In order to facilitate the search for point sources, the extended galaxy light was
subtracted.  This was done using a median filter with an inner radius of
$9.5\arcsec$ and an outer radius of $11\arcsec$. This size is large enough for not
altering the point source photometry which has been verified with artificial star
tests described later on.

The photometry has been done using the PSF fitting routine {\it allstar} within 
DAOPhot II.

\begin{table} % []
\label{tab:basicdata}
\begin{tabular}{ccccc}
\hline
\hline
Object & RA (J2000) & DEC (J2000) & vel. [km/sec] & E(B-V)\\\hline
NGC 3268 & $10^h 30^m 01^s$&$-35\degr 19\arcmin30\arcsec$&$2805\pm22$&0.11\\
NGC 3258 & $10^h 28^m 54^s$&$-35\degr 36\arcmin22\arcsec$&$2792\pm28$&0.08\\
NGC 3257 & $10^h 28^m 47^s$&$-35\degr 39\arcmin28\arcsec$&$3172\pm32$&0.08\\
NGC 3260 & $10^h 29^m 06^s$&$-35\degr 35\arcmin41\arcsec$&$2416\pm32$&0.09\\
NGC 3273 & $10^h 30^m 30^s$&$-35\degr 36\arcmin37\arcsec$&$2429\pm66$&0.10\\
NGC 3271 & $10^h 30^m 27^s$&$-35\degr 21\arcmin30\arcsec$&$3794\pm66$&0.11\\
NGC 3267 & $10^h 29^m 49^s$&$-35\degr 19\arcmin23\arcsec$&$3745\pm41$&0.10\\
NGC 3269 & $10^h 29^m 58^s$&$-35\degr 13\arcmin24\arcsec$&$3799\pm41$&0.10\\
\hline
\end{tabular}
\caption{Basic data about the largest galaxies in the field. The velocities are
	taken from the RC3 catalogue. The reddening values are taken from 
	Schlegel et al. (\cite{schlegel98}).}
\end{table}

For the point source selection we used the $\chi$ and sharpness values from the PSF
fit and found approximately 11000 point sources. The brightest non-saturated
objects have $\mathrm{T1}\approx 18$ (depending slightly on the individual MOSAIC
chip).

\subsection{Photometric Calibration}

In each of the 2 nights 4 to 5 fields, each containing about 10 standard stars from
the list of Geisler (\cite{geisler96a}), have been observed with a large coverage
of airmasses (typically from 1.0 to 1.9). It was possible to use a single
transformation, since the coefficients were indistinguishable within the uncertainties 
during these nights.

We derived the following relations between instrumental and standard magnitudes: 
\begin{eqnarray} 
 \begin{split} \nonumber
	\mathrm{T1} =
	&\mathrm{r}+(0.71\pm0.01)-(0.07\pm0.01)X_\mathrm{R}\\
		&+(0.033\pm0.003)(\mathrm{C}-\mathrm{T1})\\
	\mathrm{(C-T1)} =&(\mathrm{c}-\mathrm{r})-(0.74\pm0.02)-(0.20\pm0.01)X_\mathrm{C}\\
		&+(0.088\pm0.004)(\mathrm{C}-\mathrm{T1})
\end{split}
\end{eqnarray}

The standard deviation of the difference between instrumental and calibrated
magnitudes is 0.021 in $\mathrm{T1}$ and 0.023 in $\mathrm{C-T1}$. 

\begin{figure}[t]
 \centerline{\resizebox{\hsize}{!}{\includegraphics{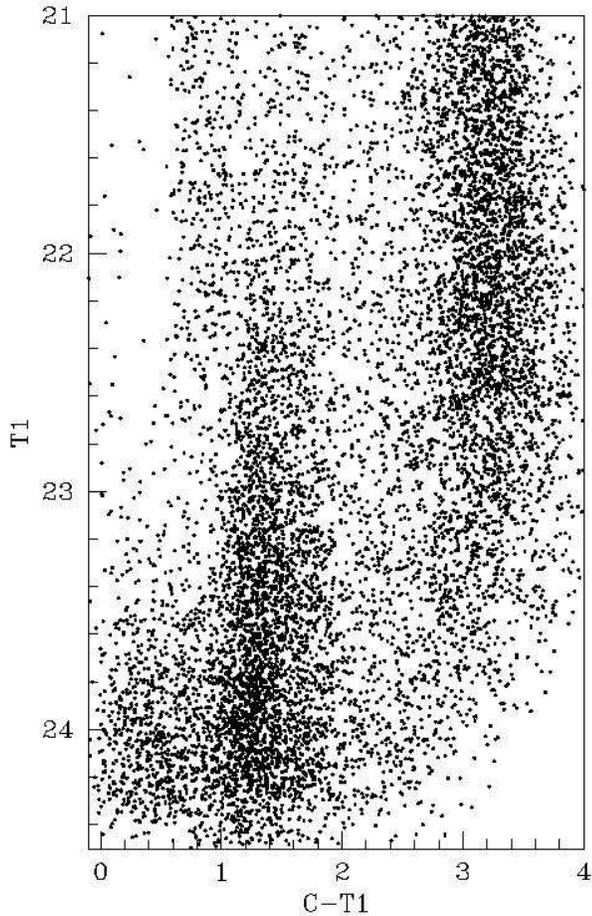}}}
 \caption{The color magnitude diagram of all point sources in the MOSAIC 
	field. Globular cluster candidates are known to occupy the color range
	of approximately $0.9<\mathrm{C-T1}<2.3$. Objects with a color
	redder than $\mathrm{C-T1}>3$ are in their majority Galactic foreground stars.
	Objects fainter than approximately T1=23.5 and bluer than $\mathrm{C-T1}=1$ 
	predominately are background galaxies.}
 \label{fig:cmdall}
\end{figure}

The color magnitude diagram for all point sources in the MOSAIC images is plotted in
Fig.\,\ref{fig:cmdall}.

The reddening towards the Antlia cluster varies according to Schlegel et al.
(\cite{schlegel98}) between E$_{\mathrm{B-V}}=0.08$ and E$_{\mathrm{B-V}}=0.11$.  
Using E$_{\mathrm{C-T1}}=1.97$\,E$_{\mathrm{B-V}}$ (Harris\,\&\,Canterna
\cite{harris77}) this means a range in reddening from
E$_{\mathrm{C-T1}}=0.16-0.22$. The IRAS map towards this direction is very patchy
and thus we decided to use just a mean reddening value of E$_{\mathrm{C-T1}}=0.19$ if not
otherwise stated (e.g. for our background comparison field).
For the absorption in R we use the relation
$\mathrm{A}_\mathrm{R}/\mathrm{A}_\mathrm{V}=0.75$ (Rieke\,\&\,Lebofsky \cite{rieke85}).

\subsection{Photometric Completeness}

\begin{figure}[t]
 \centerline{\resizebox{\hsize}{!}{\includegraphics{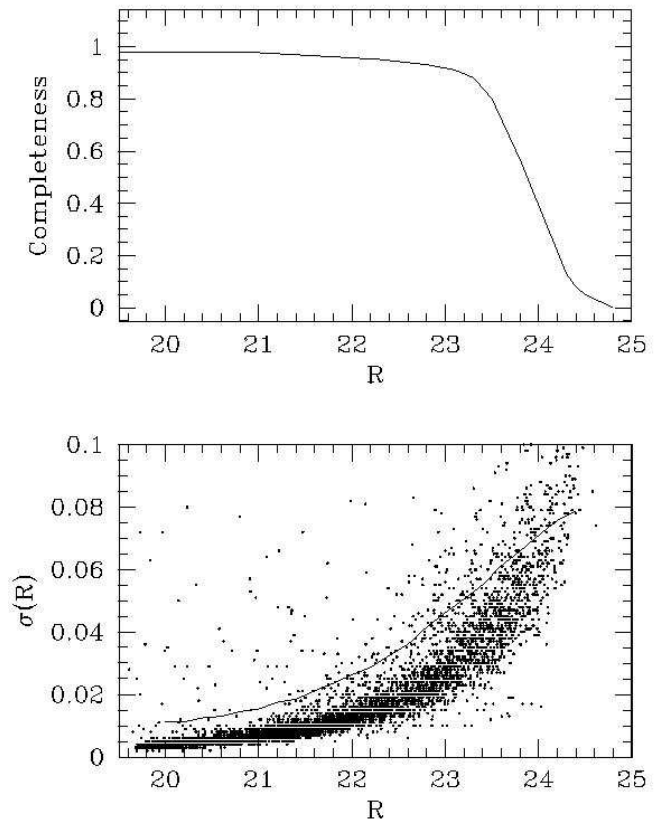}}}
 \caption{In the {\bf upper panel} the completeness function 
	for globular cluster candidates ($0.9<(\mathrm{C-T1}<2.3$) 
	is shown. In the {\bf lower panel} the DAOPhot errors (dots) are compared 
	to the mean standard deviation between the input and output magnitude of 
	the artificial stars (solid line).}
 \label{fig:compl}
\end{figure}

The completeness of the data has been studied with the aid of the task {\it
addstar} within DAOPhot II, which was used to add 6666 stars to the science image.
This was done ten times to produce ten different images. These modified images were
reduced in the same way as the original data. The final completeness function for
the color range $0.9<(\mathrm{C-T1})<2.5$ is plotted in Fig.\,\ref{fig:compl} for
the whole MOSAIC field. The difference between the completeness function for red
and blue clusters is marginal and does not need to be considered. However, there
are strong spatial variations: nearer to the center of the elliptical galaxies and
in the vicinity of bright stars the completeness is lower. Instead of using a
spatial variable completeness function we just excluded areas in which the
completeness deviates more than 5\% from its overall value.

These completeness calculations also serve to check the reliability of the
photometry and the errors given by DAOPhot. No systematic difference between input
and output magnitude were found. In the lower panel of Fig.\,\ref{fig:compl}, the
DAOPhot errors are compared to the standard deviation of the difference between
input and output magnitudes of the simulated stars.  It can be seen that this
standard deviation is larger than the DAOPhot errors and we used it instead as the
uncertainty of our measurements.

\section{Properties of the Globular Cluster Systems} 

In the Washington color $\mathrm{(C-T1)}$  globular clusters can be found in the range
$0.9<(\mathrm{C-T1})<2.3$ (e.g. Geisler et al. \cite{geisler96b}, Forte et al.
\cite{forte01}, Dirsch et al. \cite{dirsch03}). We selected as cluster candidates
all point sources within this color range that are fainter than T1=21.5. The
spatial distribution of these objects is shown in Fig.\,\ref{fig:positions}. The
GCSs of the two larger elliptical galaxies -- NGC 3258 and NGC 3268 -- can be readily
seen.

\subsection{Color Distributions}

\begin{figure}[t]
\centerline{\resizebox{\hsize}{!}{\includegraphics{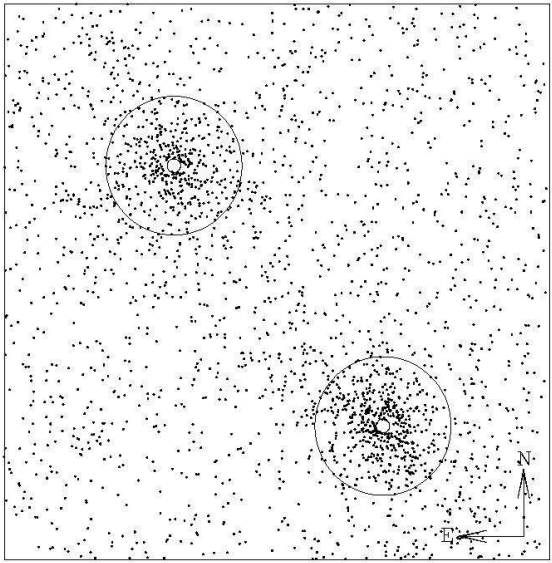}}}
\caption{The spatial distribution of all point sources fainter than T1=21.5
	within the color range $0.9<(\mathrm{C-T1})<2.3$ are plotted with 
	the same orientation as Fig.\,\ref{fig:antliafield}. We indicate the 
	areas which were used to derive the color distributions of point sources
	plotted in Fig.\,\ref{fig:colordistr_all}.}
\label{fig:positions}
\end{figure}

To derive the GC color distribution, all point sources within a radial range of
$1\arcmin <r<4\farcm5$ around NGC\,3258 and NGC\,3268 (shown in
Fig.\,\ref{fig:positions}) are selected.  Point sources within the GC candidate's
color range that are further than 10\arcmin \ away from the ellipticals and further
than 2\arcmin \ away from any other galaxy serve as a background sample. This
background might still be contaminated by globular clusters and we might slightly
underestimate the number of globulars in the galaxies

\begin{figure*}[t]
\centerline{\resizebox{15cm}{!}{\includegraphics{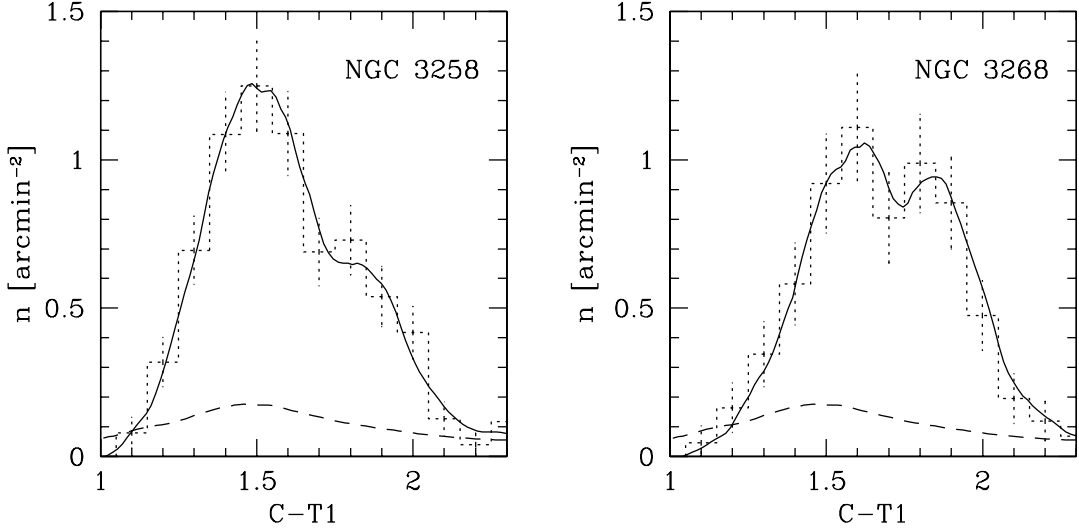}}}
\caption{The color distribution functions for the GC candidates (background 
        corrected) within a 
	radial distance of 4.5\arcmin \ around 
	NGC 3258 and NGC 3268 are shown as the solid lines.
	We used an Epanechnikov kernel for the adaptive
	smoothing. The applied background for the correction is displayed as the dashed line.
	The dotted line shows the histogram of the data and the errors.}
\label{fig:colordistr_all}
\end{figure*}

The color distributions of the background corrected cluster candidates are plotted
in Fig.\,\ref{fig:colordistr_all} where we use the
adaptive filter method described by Fadda et al.  (\cite{fadda98}) that uses an
adaptive Epanechnikov kernel (the color distribution is tabulated 
in Table\,\ref{tab:colordistrib}). This approach facilitates the visual inspection of
the distributions, in particular, it facilitates the identification of peaks and
the recognition of the peak positions. 

Around NGC\,3268 and NGC\,3258 the GCSs show up nicely as bimodal color
distributions, which is due to the good metallicity resolution of the Washington
$\mathrm{C}-\mathrm{T1}$ color. 
The two peaks can be determined by using a
phenomenological fit of two Gaussians. We applied the reddening correction 
from Table\,1 and found for the peak positions $\mathrm{C}-\mathrm{T1}=1.36\pm0.04$
and $\mathrm{C}-\mathrm{T1}=1.70\pm0.02$ for the GCS of NGC\,3258 and
$\mathrm{C}-\mathrm{T1}=1.31\pm0.02$ and $\mathrm{C}-\mathrm{T1}=1.66\pm0.03$ for
NGC\,3268. The widths of the fitted Gaussians for the blue/red peaks were 
$0.16\pm0.02$/$0.17\pm0.02$ and $0.15\pm0.02$/$0.19\pm0.02$ for
NGC\,3258 and NGC\,3268, respectively.

\begin{table}
\begin{tabular}{ccc}
\hline
\hline\\[-2.2ex]
C-T1 & $\mathrm{n}_\mathrm{NGC 3258}$ & $\mathrm{n}_\mathrm{NGC 3268}$ \\
 & $[\mathrm{arcmin}^{-2} 0.1\mathrm{mag}^{-1}$] & $[\mathrm{arcmin}^{-2}0.1\mathrm{mag}^{-1}$] \\[+0.7ex]\hline\\[-2.2ex]
1.00& $0.02\pm0.01$ & $0.00\pm0.01$\\
1.10& $0.11\pm0.03$ & $0.04\pm0.02$\\
1.15& $0.19\pm0.04$ & $0.06\pm0.03$\\
1.20& $0.33\pm0.07$ & $0.12\pm0.04$\\
1.25& $0.57\pm0.11$ & $0.21\pm0.05$\\
1.30& $0.80\pm0.14$ & $0.30\pm0.07$\\
1.35& $1.05\pm0.17$ & $0.45\pm0.10$\\
1.40& $1.22\pm0.19$ & $0.63\pm0.12$\\
1.45& $1.36\pm0.21$ & $0.86\pm0.15$\\
1.50& $1.40\pm0.22$ & $0.99\pm0.19$\\
1.55& $1.35\pm0.22$ & $1.14\pm0.22$\\
1.60& $1.20\pm0.21$ & $1.17\pm0.24$\\
1.65& $1.02\pm0.18$ & $1.09\pm0.23$\\
1.70& $0.87\pm0.16$ & $1.00\pm0.20$\\
1.75& $0.77\pm0.14$ & $0.94\pm0.18$\\
1.80& $0.73\pm0.14$ & $0.92\pm0.16$\\
1.85& $0.72\pm0.15$ & $0.97\pm0.16$\\
1.90& $0.67\pm0.15$ & $0.95\pm0.17$\\
1.95& $0.57\pm0.13$ & $0.85\pm0.16$\\
2.00& $0.40\pm0.10$ & $0.68\pm0.14$\\
2.05& $0.29\pm0.08$ & $0.54\pm0.12$\\
2.10& $0.21\pm0.06$ & $0.39\pm0.09$\\
2.15& $0.15\pm0.05$ & $0.30\pm0.07$\\
2.20& $0.11\pm0.03$ & $0.24\pm0.06$\\
2.25& $0.07\pm0.02$ & $0.17\pm0.05$\\
2.30& $0.07\pm0.02$ & $0.12\pm0.04$\\[+0.7ex]\hline
\end{tabular}
\caption{Color distribution for the cluster candidates around
	NGC 3268 and NGC 3258. A uniform reddening correction of 
	$\mathrm{C-T1}=0.19$ was applied. All
	clusters brighter than R=24 were taken into account.} 
\label{tab:colordistrib}
\end{table}

\subsection{Radial Distribution}

\begin{figure}[t]
\centerline{\resizebox{\hsize}{!}{\includegraphics{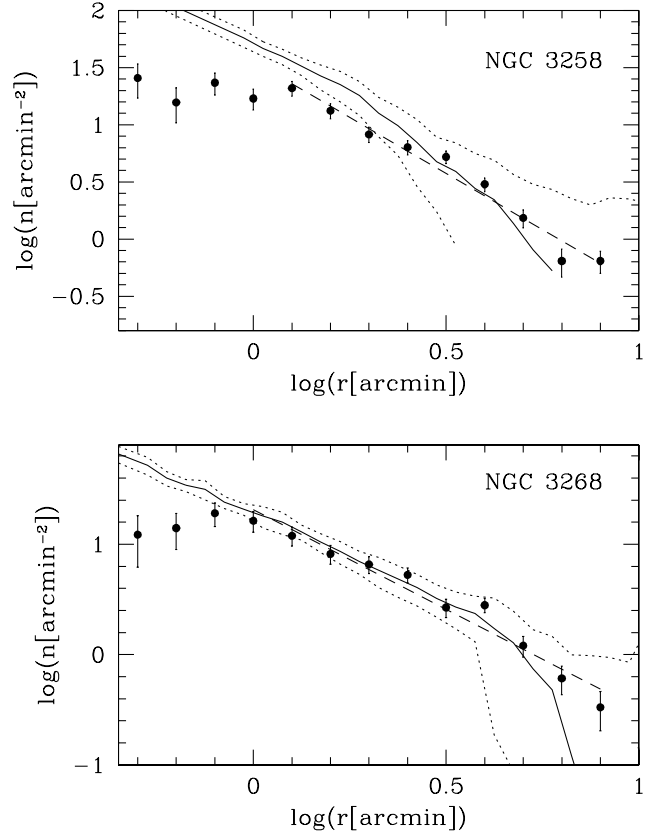}}}
\caption{This graph shows the radial distribution for the GC candidates
	($0.9<(\mathrm{C-T1})<2.3$, $21.5<\mathrm{T1}<23$) for NGC 3258 and NGC 3268.
	Within $0\farcm7$ the measurements are affected by radial incompleteness effects.
	The dashed lines show the fit to the data described in the text. The galaxy 
	T1 luminosity profile multiplied by -0.4 and arbitrarily shifted is shown
	as the solid line with $1 \sigma$ uncertainties (dotted lines). The galaxy
	measurements are described in Sect.\,4. 
	}
\label{fig:radialdistr}
\end{figure}

The radial distributions of the cluster candidates around NGC 3258 and NGC 3268 are
plotted in Fig.\,\ref{fig:radialdistr} and tabulated in Table\,\ref{tab:densitydistrib}. 
The radial number distribution of the GCS is most frequently described as a power-law 
and we find, using the clusters within 1\arcmin \ and 10\arcmin \ , a power-law index of
$-1.96\pm0.13$ for NGC\,3258 and of $-1.81\pm0.13$ for NGC\,3268. These fits are 
indicated in Fig.\,6 which shows that in particular for NGC\,3258 a single power-law is a
rather poor fit. To account for the changing slope we use a function which is motivated by
the form of a modified Hubble distribution (Binney \& Tremaine \cite{binney87}) which
includes a core radius $r_0$ and which converges to a power-law of index $2\beta$ at
large radii:
$$
n(r) = a \left(1+\left(\frac{r}{r_0}\right)^2\right)^{-\beta},
$$
where $n(r)$ is the surface number density.
We found that the following parameters provided the best fit:
\begin{itemize}
\item{NGC 3258: $a=23\pm7$, $r_0=1.7\arcmin\pm0.5\arcmin$, $\beta=1.3\pm0.2$}
\item{NGC 3268: $a=20\pm6$, $r_0=1.5\arcmin\pm0.5\arcmin$, $\beta=1.2\pm0.2$}
\end{itemize}

It is neither possible to detect any difference in the distribution of red and blue
clusters, as frequently seen in other elliptical galaxies (e.g. NGC 1399, Dirsch et
al. \cite{dirsch03}, NGC 4472, Rhode\,\&\,Zepf \cite{rhode01}), nor to separate 
the GCS into two populations with different mean colors and  different radial 
distributions. Perhaps these effects are blurred by small number statistics and deeper
observations are needed to clarify this point.

\begin{table}
\begin{tabular}{ccc}
\hline\\[-2.2ex]
$\log(\mathrm{r})$ & $\log(\mathrm{n}_\mathrm{NGC 3268})$ & $\log(\mathrm{n}_\mathrm{NGC 3258})$ \\
$[\mathrm{arcmin}]$ & $[\mathrm{arcmin}^{-2}$] & $[\mathrm{arcmin}^{-2}$] \\[+0.7ex]\hline\\[-2.2ex]
 -0.3   &$  1.41^{+0.12}_{-0.17}$&$ 1.09^{+0.17}_{-0.29}$\\
 -0.2   &$  1.20^{+0.13}_{-0.18}$&$ 1.15^{+0.13}_{-0.20}$\\
 -0.1   &$  1.37^{+0.09}_{-0.11}$&$ 1.28^{+0.09}_{-0.12}$\\
  0.0   &$  1.23^{+0.08}_{-0.10}$&$ 1.21^{+0.08}_{-0.10}$\\
  0.1   &$  1.32^{+0.06}_{-0.07}$&$ 1.08^{+0.08}_{-0.10}$\\
  0.2   &$  1.12^{+0.06}_{-0.07}$&$ 0.91^{+0.08}_{-0.09}$\\
  0.3   &$  0.92^{+0.06}_{-0.07}$&$ 0.82^{+0.07}_{-0.08}$\\
  0.4   &$  0.80^{+0.06}_{-0.07}$&$ 0.72^{+0.06}_{-0.07}$\\
  0.5   &$  0.72^{+0.05}_{-0.06}$&$ 0.43^{+0.08}_{-0.09}$\\
  0.6   &$  0.48^{+0.06}_{-0.07}$&$ 0.45^{+0.06}_{-0.07}$\\
  0.7   &$  0.18^{+0.07}_{-0.09}$&$ 0.08^{+0.08}_{-0.10}$\\
  0.8   &$ -0.19^{+0.11}_{-0.14}$&$-0.22^{+0.11}_{-0.15}$\\
  0.9   &$ -0.19^{+0.09}_{-0.11}$&$-0.48^{+0.14}_{-0.21}$\\[+0.7ex]\hline
\end{tabular}
\caption{Radial cluster density distribution ($0.9<(\mathrm{C-T1})<2.3$, 
	$21.5<\mathrm{T1}<23$) for NGC 3268 and NGC 3258. 
	}
\label{tab:densitydistrib}
\end{table}

\subsection{Azimuthal Distribution}

\begin{figure}[t]
\centerline{\resizebox{\hsize}{!}{\includegraphics{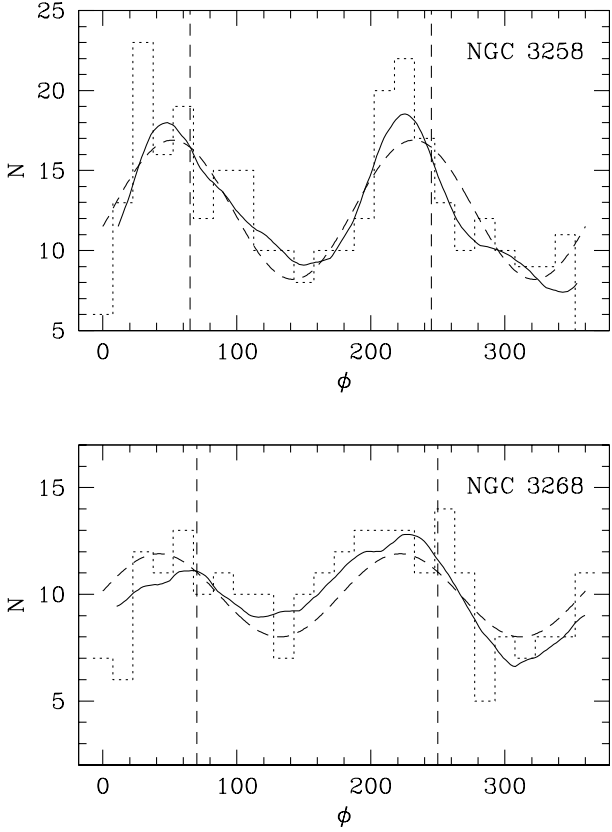}}}
\caption{The azimuthal distribution of the globular clusters within a radial
	distance of $1\farcm6<r<4\farcm5$ around NGC\,3258 and NGC\,3268.
	The dotted line shows the binned azimuthal distribution and the 
	solid line the smoothed one that was also used for a $\sin^2$-fit shown
	as the dashed line. This graph shows that both GCSs have an elliptical shape.
	The vertical dashed lines show the position angle of the galaxy light 
	(see Sect.\,4).}
\label{fig:azi}
\end{figure}

The azimuthal number densities of the GCSs of NGC\,3258 and NGC\,3268 
within a radial range
$1\farcm6<r<4\farcm5$ are shown in Fig.\,\ref{fig:azi}. The azimuthal angle is 
defined as a position angle, measured from North to East. 
An elliptical GCS causes sinusoidal counts in this diagram with the ellipticity
$\epsilon=1-(N_b/N_a)^{1/\alpha}$ ($N_b,N_a$ being the number of clusters along
minor and major axis respectively and $\alpha$ the absolute value of the exponent
of the radial distribution).

The GCS of NGC\,3258 is clearly elongated and a sinusoidal fit (also shown in
Fig.\,\ref{fig:azi}) yielded an ellipticity of
$\epsilon=0.26\pm0.09$ and a position angle of $38\degr\pm6\degr$. Also the GCS of
NGC\,3268 is elongated, but it has a lower ellipticity of $\epsilon=0.15\pm0.09$.
Its position angle has been determined to be at $47\degr\pm9\degr$. 

\subsection{The Total Number of GCs}

The total number of clusters can of course not be measured directly, yet it is well
known that the luminosity function of GCs can be described (in magnitudes)
with a Gaussian that has a characteristic turn-over magnitude (TOM) and a limited
range of possible widths. Assuming a distance towards the two elliptical galaxies,
a TOM and a width of the Gaussian, we can extrapolate the number of observed
GCs to fainter magnitudes covering the whole luminosity function. We assumed a TOM
for the Milky Way GCS of $\mathrm{V}=-7.61\pm0.08$ and a metallicity correction
derived by Ashman et al. (\cite{ashman95}) of 0.2 (corresponding to a mean cluster
metallicity of -0.6\,dex, adequate for the two elliptical galaxies). As we do not have
V observations, we have to assume a $\mathrm{V-R}$ color of $\mathrm{V-R}=0.6$ (taken from the 
Milky Way clusters of a similar metallicity). Thus, we expect the TOM to be at
$\mathrm{M}_\mathrm{TOM} = -8.0$ in R and hence an unreddened $\mathrm{m}_\mathrm{TOM} =
24.7\pm0.25$. This value can also be used for T1 since there is only a negligible
zero point difference between T1 and R with very little color dependence. For
the width of the Gaussian, values from 1.2 to 1.4 are typically found in elliptical
galaxies of this size in the literature (Ashman\,\&\,Zepf \cite{ashman98} and
references therein).

\begin{figure}[t]
\centerline{\resizebox{\hsize}{!}{\includegraphics{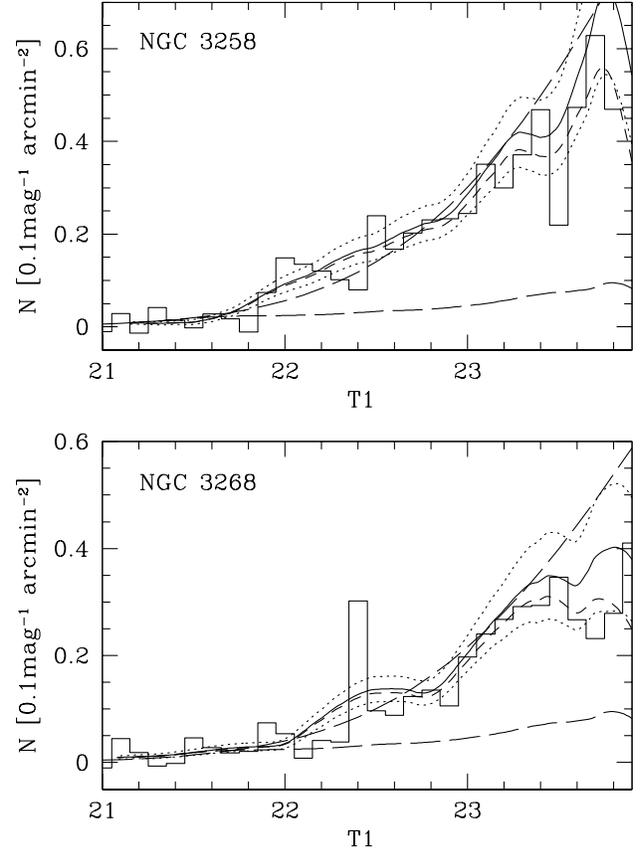}}}
\caption{The completeness corrected luminosity function of GC candidates 
	($1<(\mathrm{C-T1})<2$) with a radial distance between 
	$0.46\arcmin<\mathrm{r}<4.5\arcmin$ is plotted as histogram and 
	adaptively  filtered distribution with a solid line for NGC 3258 
	in the {\bf upper panel} and for NGC\,3268 in the {\bf lower panel}. 
	The short dashed line shows the luminosity function before the
	completeness correction. The lower long dashed lines in the panels 
	indicate the background luminosity function and the upper long dashed 
	line the fitted Gaussian described in the text.} 
\label{fig:lumfkt}
\end{figure}

We used clusters within a radial distance of 4.5\arcmin \ to construct the globular
cluster luminosity functions (GCLF) that are shown in Fig.\,\ref{fig:lumfkt}. The
GCLF of NGC 3268 is peculiar: it shows an enhancement of point sources around
$22.4<\mathrm{T1}<22.6$. These objects are not spatially concentrated in any particular way.
Also the apparent deficiency of point sources fainter than
$\mathrm{T1}=23.4$ is striking, since we cannot have reached the TOM at such bright
magnitudes. The reason might be an underestimation of the completeness at these
magnitudes, since in the vicinity of NGC\,3268 there are 4 brighter spiral galaxies. 
However, in the following we take only clusters brighter than 23.5 into account and
considering the other uncertainties mentioned below, we do not regard this as a
severe problem. Further observations are nevertheless required for a deeper and
more complete
study of this peculiar luminosity function.

Since we observe only $\approx 10\%$ of all clusters around the galaxies,
changes in the distance and the width of the Gaussian can result in large uncertainties 
in the total number of GCs.
We find for the total
number of GCs, within a radial distance of $4\farcm5$, around NGC\,3258:
$3400\pm280^{+3600+900}_{-1200-600}$, and around NGC\,3268:
$2500\pm220^{+2600+800}_{-900-500}$. The first uncertainty is the statistical
uncertainty, the second one is due to the distance uncertainty, and the third 
one is due to the
uncertainty in the width of the distribution. The large uncertainties due to the
distance uncertainty and the width of the distribution, do not allow a firmer 
statement than that the
GCSs of both galaxies encompass several thousands GCs and that NGC\,3258
contains more clusters within $4\farcm5$ (corresponding to $5r_\mathrm{eff}$ and
$3.3r_\mathrm{eff}$ for NGC\,3258 and NGC\,3268, respectively) than NGC\,3268.

\section{Properties of the Host Galaxies}

\subsection{Galaxy Luminosity Profiles}

We determined the luminosity profiles of NGC\,3258 and NGC\,3268 with the purpose
of comparing them with their GCSs. The longer exposures were used for the outer part 
(radial distances larger than 30\arcsec) and the short one for the inner part. 
Within $2\farcm5$, the task {\it
ellipse} within IRAF can be used to determine the profile; outside, however, the
results obtained with {\it ellipse} were not satisfactory because of the many
disturbing objects and thus we decided to use a similar approach as in Dirsch et
al. (\cite{dirsch03}): we measured the luminosity in 3\,pix apertures in 33333
randomly distributed areas around NGC\,3258 and NGC\,3268 on the point source
subtracted images. We then excluded areas that are influenced by galaxies, strong
stellar residuals and saturated stars. The disadvantage of this approach is that it
is not straightforward to determine the ellipticity and thus this has been done
only in the inner $2\farcm5$.

\begin{figure}[t]
\centerline{\resizebox{\hsize}{!}{\includegraphics{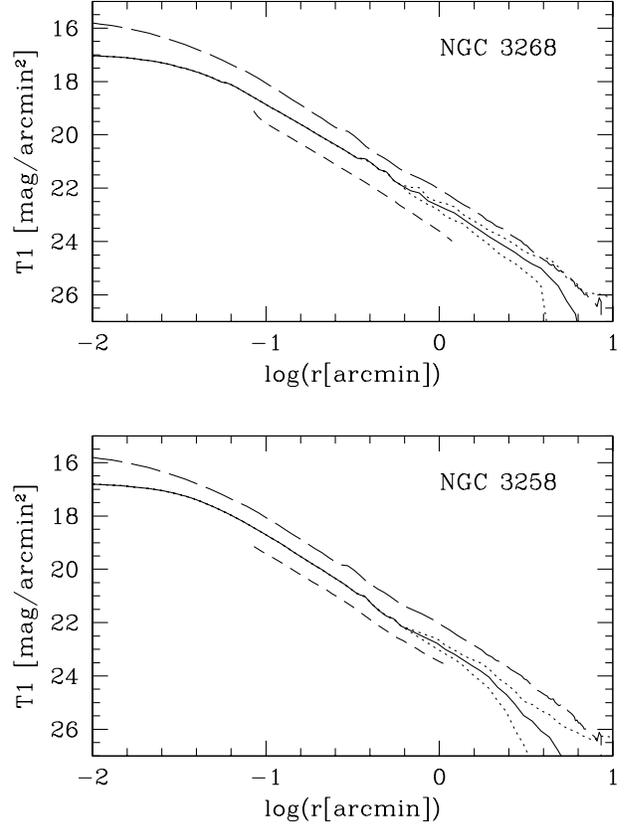}}}
\caption{The luminosity profiles of NGC\,3268 and NGC\,3258 (solid curves) and
	$1 \sigma$ uncertainties (dotted line) are compared
	to the luminosity profile of NGC\,1399 (long dashed line).
	For this comparison the diameter of NGC\,1399 has been 
	shrunk by a factor of three to account for the distance difference
	(a distance modulus of $\mathrm{(m-M)}=31.4$ is assumed for NGC\,1399).
	The short dashed lines are the V profiles for NGC\,3268 and NGC\,3258 
        published by Reid et al. (\cite{reid94}).}
\label{fig:lum_profile}
\end{figure}

The luminosity profiles are plotted in Fig.\,\ref{fig:lum_profile}. For comparison,
we included in this figure the luminosity profiles of the central Fornax cluster
galaxy NGC\,1399 (taken from Dirsch et al. \cite{dirsch03}).  For this comparison
we shifted the Fornax galaxy to the Antlia distance.  In the inner part
($r<3.5\arcsec$), the profiles of the Antlia galaxies appear flatter than that of
NGC\,1399. The reason is that HST observations of Lauer et al. (\cite{lauer95})  
were used to determine the luminosity profile in the inner part of NGC\,1399,
where the PSF is much smaller than for our data. In
Fig.\,\ref{fig:lum_profile} we included the V profiles determined by Reid et al.
(\cite{reid94}) for NGC\,3268 and NGC\,3258. The agreement is excellent and a 
mean $\mathrm{V}-\mathrm{R}$ color of
$\mathrm{V}-\mathrm{R}\approx0.7$ can be determined. This color agrees with the
value given by Godwin et al. (\cite{godwin77}) for NGC\,3268.

The luminosity profiles for radii smaller than $5\arcmin$ can be fitted with
the same function as used for the cluster density distribution:
$$
\mathrm{T1}=a \log\left(1+(r/r_0)^2\right)+\mathrm{T1}_0
$$
The following values have been found:
\begin{itemize}
\item{NGC 3258: $a=1.99\pm0.01$, $r_0=0.036\arcmin\pm0.001\arcmin$, $\mathrm{T1}_0=16.77\pm0.01$}
\item{NGC 3268: $a=2.34\pm0.02$, $r_0=0.044\arcmin\pm0.001\arcmin$, $\mathrm{T1}_0=16.57\pm0.03$}
\end{itemize}

\begin{figure}[t]
\centerline{\resizebox{\hsize}{!}{\includegraphics{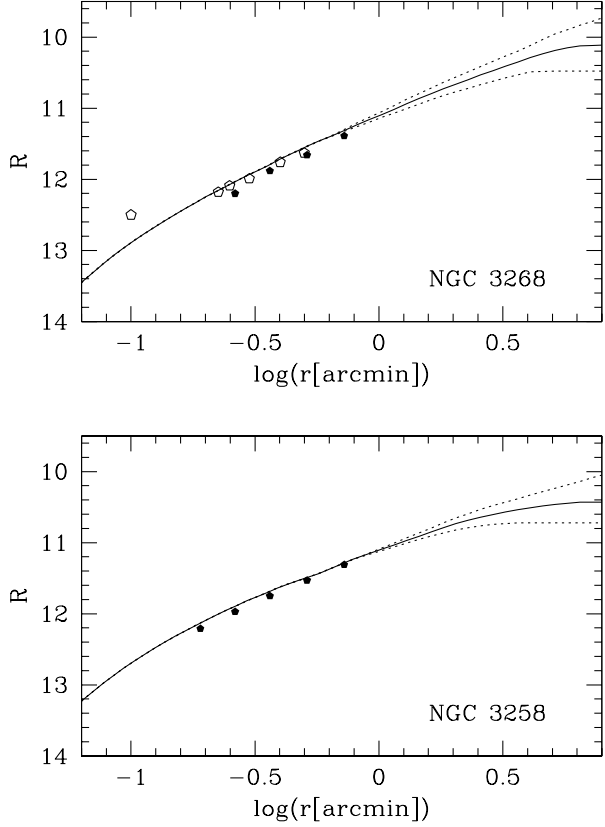}}}
\caption{The integrated luminosity dependence on radius is shown for NGC\,3268
	and NGC\,3258 in the R band (we tranformed our T1 measurements using the
	tiny shift mentioned in Sect.\,2). For NGC\,3268 the 
	measurements of Godwin et al. (\cite{godwin77})
	are plotted as open symbols. For NGC\,3268 and NGC\,3258 aperture photometry
	is provided by Poulain et al. (\cite{poulain94}) (filled symbols).}
\label{fig:lum_profile_integ}
\end{figure}

In Fig.\,\ref{fig:lum_profile_integ} the integrated luminosity in R is shown
(we transformed our T1 measurements into R using the tiny shift mentioned in Sect.\,2).
Poulain et al. (\cite{poulain94}) determined the luminosity of NGC\,3258 within five 
apertures and their results are shown in Fig.\,\ref{fig:lum_profile_integ} as well.
For NGC\,3268, aperture photometry has been performed by Godwin et al.
(\cite{godwin77}) and Poulain et al. (\cite{poulain94}), and their measurements are 
also displayed in Fig.\,\ref{fig:lum_profile_integ}.

By means of the R integrated luminosity profiles it is also possible to 
perform a rough estimation of the effective radii ($r_\mathrm{eff}$) of NGC\,3258 and 
NGC\,3268, assuming that the values obtained at 
the radius of 5\arcmin \ (about 50\,kpc) represent the total integrated luminosities. 
Thus, the measured effective radii result are $(8 \pm 4)$\,kpc for NGC\,3258 and 
$(12 \pm 5)$\,kpc for NGC\,3268, where the large errors are a consequence of 
the uncertainties in the luminosity profiles at large radii. 

\subsection{Galaxy Color Profile}

\begin{figure}[t]
\centerline{\resizebox{\hsize}{!}{\includegraphics{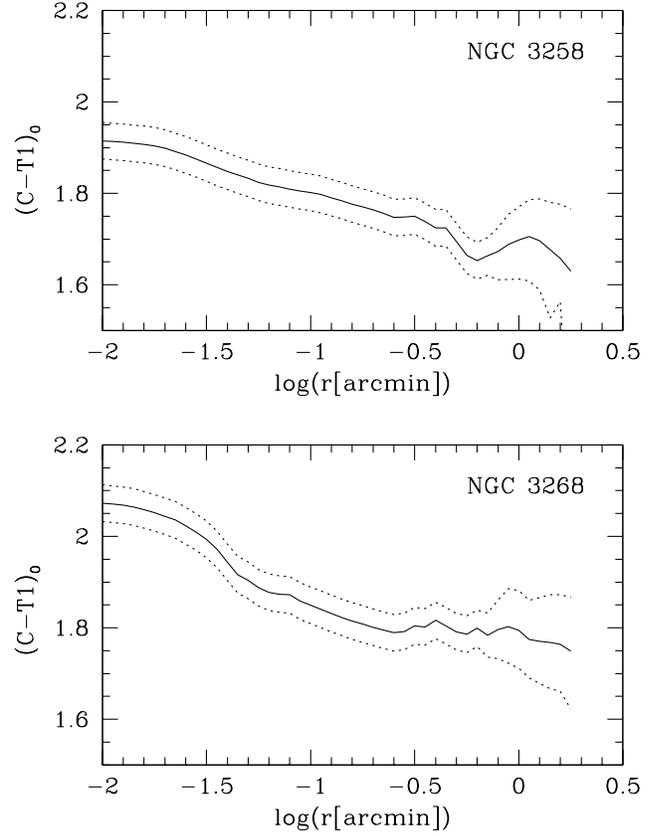}}}
\caption{The color profiles of the two ellipticals are shown together
	with the $1 \sigma$ uncertainties. }
\label{fig:colorprofile}
\end{figure}

The C luminosity profiles are used together with 
the T1 profiles to search for possible color gradients, which is shown in 
Fig.\ref{fig:colorprofile} for the dereddened colors. In this Figure it is 
apparent that both galaxies become bluer with increasing radius. 
However, the color profiles of both galaxies are quite different at 
radial distances smaller than 3\arcsec. Within this radius NGC\,3268 becomes remarkably red.
Visual inspection of the C image showed a clearly off-centered central peak. 
Both observations can be explained by an inclined circumnuclear dusty disk with a radius 
of roughly 3\arcsec (which corresponds to 500\,pc).

The observed color gradient of NGC\,3258 is in accordance with the B-V gradient
found by  Reid et al. (\cite{reid94}). However, they found no
color gradient in NGC 3268. This might be explained by the fact that the
color gradient of this galaxy is not
as pronounced as in NGC\,3258 at larger radii.
In (V-I) no color gradient is apparent for either galaxy (Reid et al. \cite{reid94}).

\subsection{Galaxy Ellipticities}

The ellipticities have been measured within the inner $2\farcm5$ using the
task {\it ellipse}. For NGC\,3268, the derived ellipticity  
agrees well within 1\arcmin \ with the values derived by Reid et al. (\cite{reid94}); 
further outside, however, the ellipticity of Reid et al. increases while in our model 
it remains constant at $\epsilon=0.20\pm0.08$ out to
$2\farcm5$. The position angle of $\approx70\degr$ of Reid et al.
(\cite{reid94}) agrees with our value of $70\degr\pm5$ up to $2\farcm5$ (the
uncertainties were derived by varying initial conditions and fit parameters of the
task {\it ellipse}).
Also for NGC\,3258, our results confirm the earlier measurements by Reid 
et al. (\cite{reid94}): the ellipticity
increases from 0.1 to about 0.3 at $2\farcm5$. The position angle decreases in the
same radial range from $75\degr\pm5\degr$ to $65\degr\pm5\degr$.

The azimuthal distributions of the GCSs, presented in Sect.\,3.3, are estimated 
within the radial range $1\farcm6<r<4\farcm5$,  that is, partially overlapping the area 
over which the galaxies' ellipticities have been calculated ($r<2\farcm5$). We 
cannot strictly compare the GCS and the galaxy light within the same radial range, since
at small radii the incompleteness does not allow reliable determination of the GCS 
ellipticity and at radii larger than $2\farcm5$ the ellipticity of the galaxy cannot 
be determined due to the low surface brightness and the accompanying galaxies.

\subsection{Comparing the light profiles of NGC\,3268/3258 and NGC\,1399}

It is interesting to compare the luminosity profiles of NGC\,3268 and NGC\,3258
with the profile of the central Fornax cluster galaxy 
NGC\,1399 (Fig.\,\ref{fig:lum_profile}). This
comparison shows that the luminosity profile of NGC\,3268 resembles
that of NGC\,1399 very closely. The luminosity profile of NGC\,3258
seems to be steeper than that of NGC\,1399 at radii larger than
2\arcmin \ . However, within the uncertainties it is also consistent 
with having no change in the slope. Better data is needed for an interpretation 
of its profile.

The surface luminosity of both
galaxies is considerably lower that that of NGC\,1399 (projected to the same
distance), the reason being foreground reddening which accounts for an
absorption in R of 0.26 for NGC\,3268 and 0.19 for NGC\,3258. 

The deprojected luminosity density of NGC\,3268 thus is close
to a $r^{-3}$ profile as in the case of NGC\,1399 (Dirsch et al.
\cite{dirsch03}) with a decline to a steeper profile only at very large radii.
NGC\,3258 shows a steeper profile at clearly smaller radii.
The transition from a $r^{-3}$ to a $r^{-4}$ density profile, which in projection
gives the familiar de Vaucouleurs profile, is probably caused by a sharp cut in the energy
distribution function near the escape energy in the case of an approximate
Keplerian potential, as has been argued
by Jaffe (\cite{jaffe87}) and White (\cite{white87}). In this line of argument Dirsch et al.
(\cite{dirsch03})
attributed the $r^{-3}$ decline (deprojected) of NGC 1399 to the presence
of its strong dark matter potential.  In analogy  also 
NGC\,3268 may be located  inside a deep potential  of 
dark matter that constitutes the Antlia cluster, which is consistent with the
results from the X-ray observations.

\section{Comparing the GCS and the Galaxy Light}

\subsection{Specific Frequency}

The specific frequency is the number of clusters (N)  
normalized to the V luminosity of the galaxy ($\mathrm{S}_\mathrm{N}=\mathrm{N}\cdot
10^{0.4(\mathrm{M}_\mathrm{V}+15)}$). Hence the total luminosity of the galaxy is 
required.

In our case the large luminosity uncertainties at
8\arcmin , our outermost radius, do not permit us to derive a reliable global
specific frequency. To circumvent this problem we consider only the circular area
with a radius of 4\arcmin \ for the calculation of the specific frequency, which 
corresponds to about 40\,kpc at the distance of the Antlia cluster, or approximately 
$5\,r_\mathrm{eff}$ for NGC\,3258 and $3.3\, r_\mathrm{eff}$ for NGC\,3268, according to the 
rough values obtained in Sect.\,4.1. This is the largest radius for which a reasonably
accurate total luminosity can be obtained (see Fig.\,10). The
large uncertainty of the absolute magnitude and of the total number of clusters due to
the distance uncertainty is not too critical, since these uncertainties are
coupled - i.e. a larger distance means a brighter galaxy, but also more globular
clusters. We used a color of $\mathrm{V-R}=0.7$ for NGC\,3268 (Godwin et al.
\cite{godwin77}, slightly redder than the color used for the GCs above,
$\mathrm{V-R}=0.6$). For the reddening correction the relation given by
Rieke\,\&\,Lebofsky (\cite{rieke85}), $\mathrm{E(V-R)}/\mathrm{E(B-V)}=0.78$ is
used. The luminosity within 4\arcmin \ is $\mathrm{R}=10.22\pm0.15$ and
$\mathrm{R}=10.45\pm0.18$ for NGC\,3268 and NGC\,3258, respectively. Using our 
distance modulus and the reddening values from Table\,1 we obtain for the
absolute luminosity of NGC\,3268 and NGC\,3258, $\mathrm{M}_\mathrm{R}=-22.77\pm0.30$ and
$\mathrm{M}_\mathrm{R}=-22.10\pm0.31$, respectively. With $\mathrm{V-R}=0.7$,
we finally
find a specific frequency of $\mathrm{S}_\mathrm{N}=3.0\pm2.0$ for NGC\,3268 and of
$\mathrm{S}_\mathrm{N}=6.0\pm2.6$ for NGC\,3258 (the mean value is calculated for a width of
the Gaussian GCLF of 1.3 and a distance of $\mathrm{(m-M)}=32.73$). 

\subsection{Morphological Comparison}

In Fig.\,\ref{fig:radialdistr} the luminosity profiles are compared to the radial cluster
distributions. 
For NGC\,3268, both radial profiles agree very well for radii larger
than $0\farcm7$. At smaller radii the clusters have a shallower distribution, which 
predominately is caused by the radial incompleteness variations. When 
only brighter clusters are used, the slope in the inner region increases; however, 
the noise also increases due to the smaller number statistics. In the outer region, the 
slope remains
independent from the choice of the lower limiting magnitude within the uncertainties.
In NGC\,3258 the large uncertainties of the luminosity profile in the overlapping area 
make a statement more difficult.
Within the uncertainties, the GC density distribution and luminosity profile agree,
albeit there is an indication that the clusters have a slightly shallower
distribution than the T1 profile. Such a behavior is expected in a galaxy in 
which a metal poor cluster population has a shallower distribution than a more metal rich 
one, since the galaxy luminosity profile is measured in a relatively red passband. 
Since clusters are accompanied by field populations, the different distribution of the
two cluster populations should be reflected in
the field populations. This picture agrees well with the fact that NGC\,3258
shows a color gradient. Nevertheless, as stated above, we were unable to disentangle the GCS into
two populations that have a different mean color and a different radial distribution. 
However, this might be due to small number statistics and further observations are required 
to clarify this strange behavior.

The cluster systems and the galaxies are elliptical with similar ellipticities at
overlapping radii. NGC\,3258 becomes progressively more elliptical with radius, 
which is also reflected in its rather elongated GCS, while NGC\,3268 maintains
approximately its shape. The position angles deviate slightly between galaxy light and
cluster system in both galaxies. This deviation might, however, be understandable for NGC\,3258
where the position angle of the galaxy light
monotonically changes with radius and, extrapolating it to
the GCS radial regime, we expect the observed GCS position angle also for the galaxy
light.

It is noteworthy that the position angles of the GCS in both galaxies are
approximately aligned with the connecting line of the two galaxies that has a
position angle of $50\degr$.

\section{Summary}

This investigation demonstrated
the presence of a few thousand GCs in NGC\,3258 and NGC\,3268. The resulting 
specific frequencies are $\mathrm{S}_\mathrm{N}=3.0\pm2.0$ for NGC\,3268 and 
$\mathrm{S}_\mathrm{N}=6.0\pm2.6$ for NGC\,3258. These specific frequencies 
were calculated within a radius of 40 kpc, which corresponds to 
approximately $5\,r_\mathrm{eff}$ for NGC\,3258 and $3.3\, r_\mathrm{eff}$ for NGC\,3268, 
as said above.  
The specific frequency of NGC\,3268 is typical for
an elliptical galaxy (e.g. Ashman\,\&\,Zepf \cite{ashman98}), while the one of NGC 3258 
is comparable to giant ellipticals in the center of galaxy clusters, like
NGC 1399 (Dirsch et al. \cite{dirsch03}). However, given the
uncertainties, these values should be regarded as being indicative only.

The color of the red and blue peaks of the GCS 
appear to be very similar to that of other galaxies that
were studied in the Washington system (NGC\,1399 - Dirsch et al. \cite{dirsch03}, 
NGC\,4472 - Geisler et al. \cite{geisler96b}, M\,87 - C\^ot\'e et al. \cite{cote01},
NGC\,1427 - Forte et al. \cite{forte01}). This was expected given that 
investigations in V and I also resulted in similar, if not equal colors
(e.g. Kundu \& Whitmore \cite{kundu01a}, Forbes\,\&\,Forte \cite{forbes01}, 
Larsen et al. \cite{larsen01}, Gebhardt\,\&\,Kissler-Patig \cite{gebhardt99}).

The luminosity profile of NGC\,3268 resembles that of NGC\,1399 
remarkably well.

We compared the spatial distribution of the stellar light and the GCS and found
(at least in the outer range -- $r>1.7\arcmin$) the same profiles. 
Differences in the distribution of red and blue subpopulations are not discernible
due to poor number statistics.
Furthermore, the major axes of both galaxies are 
within the uncertainties aligned with their connecting axis, resembling the X-ray 
findings. 
Whether this is due to a physical process or whether it
is a chance alignment can not be answered by this study.

\appendix

\section{A strange absorption feature in NGC 3269}

\begin{figure}[t]
\centerline{\resizebox{\hsize}{!}{\includegraphics{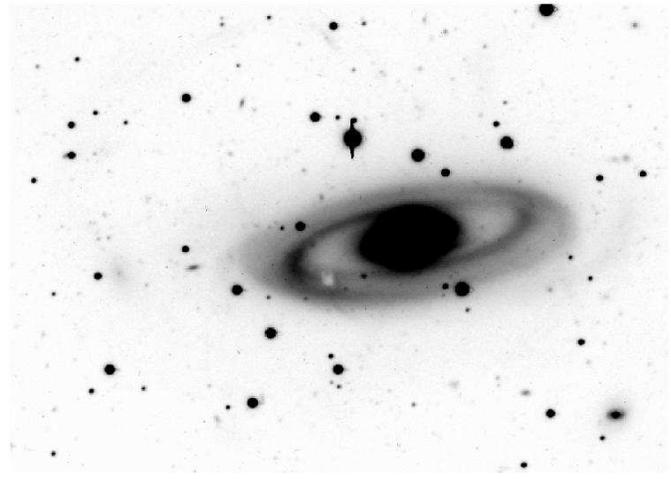}}}
\caption{C image of NGC 3269 that shows the compact absorption feature in the
	spiral arm to the lower left. North is to the right, East to the top. The size
	of the image is $4.9\arcmin\times3.7\arcmin$.}
\label{fig:n3269C}
\end{figure}

During the visual inspection of the frames we noted a compact absorption feature in
NGC 3269, a spiral galaxy that has no noticeable GCS. The dust cloud can be seen in
the south-western part as shown in a C image (Fig.\,\ref{fig:n3269C}). NGC 3269 is strikingly devoid of gas (Barnes et al.
\cite{barnes01}) and exhibits a very smooth structure. The absorption feature has a
diameter of 4\arcsec and can also been seen on previous images of NGC\,3269, for
example in the Sandage\,\&\,Betke (1994) atlas. At the distance of NGC 3269 it would have a
diameter of approximately 500\,pc. It is an intriguing question whether this cloud
is indeed in NGC\,3269 or whether it is a foreground object in the Galaxy. One
approach for further insight is to study its reddening law.

One can estimate the reddening law assuming that the obscured part is similar to
the adjacent surface luminosity of the galaxy, which is a reasonable assumption
considering its smooth appearance. We obtained by linear interpolation between 
adjacent regions an
absorption of $\mathrm{A}_\mathrm{C}=0.62\pm0.08$ and $\mathrm{A}_\mathrm{R}=0.29\pm0.08$. 
In analogy to the absorption in V, we define
$\mathrm{R}_{R}=\frac{\mathrm{A}_\mathrm{R}}{\mathrm{E}_\mathrm{C-R}}=0.9\pm0.3$. Using
the relations of Harris\,\&\,Canterna (\cite{harris77}) and
$\mathrm{R}_{V}=\frac{\mathrm{A}_\mathrm{V}}{\mathrm{E}_\mathrm{B-V}}=3.2$ we would
expect
$\mathrm{R}_\mathrm{R}=\frac{0.75\mathrm{A}_\mathrm{V}}{1.97\mathrm{E}_\mathrm{B-V}}=1.22$.
The main result is that the derived R value is not {\it larger} than the expected value which 
would imply a grayer reddening law.

The singular existence of a dusty region of 500\,pc diameter in a galaxy which
otherwise is devoid of gas and dust is by itself peculiar. Moreover, it should have
substructure leading to small-scale opacity variations for different lines of
sight. Unless it is located well above the plane of NGC\,3269 (which would be
another peculiar feature) we expect light contribution from the foreground stellar
population.  Both facts would result in a grayer reddening
law, i.e. a higher R value. Thus the observation of a rather lower R value is an
indication that this cloud indeed is located within our galaxy. To give an idea
of the possible cloud size in this case, we might
estimate an upper limit of its size by assuming a scale height of the
galactic disk of 800\,pc (which is certainly rather large); 
the diameter would be less than 0.06\,pc.
Dust clouds with these properties should be extremely difficult to detect, unless
they are by chance projected onto a suitable background source.

It could be the optical detection of the same type of compact Galactic gas clouds reported
by Heithausen (\cite{heithausen02}). He found two molecular structures smaller than
1\arcmin \ (one being $\le 25\arcsec$ in CO). The absorption was estimated to be
$\mathrm{A}_\mathrm{V}\le0.2$ via the H{\sc i} column density, which is of the same
order as our determined absorption ($\mathrm{A}_\mathrm{V}=0.4$).

\section*{Acknowledgments}
TR and BD gratefully acknowledge support from the Chilean
Center for Astrophysics FONDAP No. 15010003. LB is also thankful to the Astronomy 
Group in Concepcion for the financial support and warm hospitality.
BD gratefully acknowledges financial support of the Alexander von Humboldt
Foundation via a Feodor Lynen Stipendium and spiritual support in the ''30 y
tantos''.

\end{document}